\documentclass[english]{IEEEtran}
\usepackage[T1]{fontenc}
\usepackage[latin9]{inputenc}
\usepackage{color}
\usepackage{amsmath}
\usepackage{amssymb}
\usepackage{stmaryrd}
\usepackage{graphicx}

\makeatletter
\author{
    \IEEEauthorblockN{Abubakr O. Al-Abbasi\IEEEauthorrefmark{1}, Ridha Hamila\IEEEauthorrefmark{1}, Waheed U. Bajwa\IEEEauthorrefmark{2}, and Naofal Al-Dhahir\IEEEauthorrefmark{3}}

    \IEEEauthorblockA{\IEEEauthorrefmark{1}
Dept. of Electrical Engineering, Qatar University, Qatar
}

\IEEEauthorblockA{\IEEEauthorrefmark{2}
Dept. of Electrical and Computer Engineering, Rutgers University, USA
}

    \IEEEauthorblockA{\IEEEauthorrefmark{3}
Dept. of Electrical Engineering, University of Texas at Dallas, USA
}

\thanks{This paper was made possible by grant number NPRP 06-070-2-024 from
the Qatar National Research Fund (a member of Qatar Foundation). The
statements made herein are solely the responsibility of the authors.}
}

\usepackage{balance} 

\makeatother

\usepackage{babel}
\begin{document}

\title{Design and Analysis Framework for Sparse FIR Channel Shortening }
\maketitle
\begin{abstract}
A major performance and complexity limitation in broadband communications
is the long channel delay spread which results in a highly-frequency-selective
channel frequency response. Channel shortening equalizers (CSEs) are
used to ensure that the cascade of a long channel impulse response
(CIR) and the CSE is approximately equivalent to a target impulse
response (TIR) with much shorter delay spread. In this paper, we propose
a general framework that transforms the problems of design of sparse
CSE and TIR finite impulse response (FIR) filters into the problem
of sparsest-approximation of a vector in different dictionaries. In
addition, we compare several choices of sparsifying dictionaries under
this framework. Furthermore, the worst-case coherence of these dictionaries,
which determines their sparsifying effectiveness, are analytically
and/or numerically evaluated. Finally, the usefulness of the proposed
framework for the design of sparse CSE and TIR filters is validated
through numerical experiments.
\end{abstract}

\section{Introduction \label{sec:Introduction}}

In many applications, the channel delay spread, defined as the duration
in time, or samples, over which the channel impulse response (CIR)
has significant energy, is too long and results in performance and
complexity limitations for communications transceivers. For instance,
a large delay spread exceeding the cyclic prefix (CP) length causes
inter-symbol interference (ISI) and inter-carrier interference (ICI)
in multi-carrier modulation (MCM) systems \cite{mcm90} and increases
the complexity of sequence estimators such as the Viterbi algorithm
\cite{VAsparse98} whose computational complexity increases exponentially
with the number of CIR taps. In MCM, ISI and ICI are prevented by
inserting a CP whose length must be greater than or equal to the CIR
length, which results in a data rate reduction \cite{cyclic010}.

The aim of the channel shortening equalizers (CSEs) is to ensure that
the combined impulse response of the channel and the CSE is approximately
equivalent to a short target impulse response (TIR). Several CSE design
approaches have been investigated in the literature. In \cite{effcompRed},
a unified framework for computing the optimum settings of a TIR is
proposed under the unit-tap constraint (UTC) and the unit-energy constraint
(UEC). This framework is extended in \cite{mimoCS} to accommodate
multiple-input multiple-output (MIMO) systems. A CSE is designed in
\cite{IRS96} to maximize the output signal-to-interference ratio,
also known as the channel-shortening SNR. In \cite{falconer1973adaptive},
the mean-square error (MSE) between the CIR-CSE cascade and the TIR
is minimized subject to a UEC on the TIR. However, none of these designs
impose a sparsity constraint on the CSE to reduce its implementation
complexity. One possible approach to design a sparse filter is the
exhaustive search method \cite{chopra2012design}. However, it is
applicable only to the design of low-order sparse finite impulse response
(FIR) filters since its computational cost increases exponentially
with the filter order. Channel shortening in \cite{blindEq003} is
performed blindly in which the CSE coefficients can be inferred directly
from the received data without the channel knowledge. In \cite{newDFW},
a framework for designing sparse CSE and TIR is proposed. Using greedy
algorithms, the proposed framework achieved better performance by
designing the TIR taps to be non-contiguous compared to the approach
in \cite{effcompRed}, where the TIR taps are assumed to be contiguous.
However, this approach involves inversion of large matrices and Cholesky
factorization, whose computational cost could be large for channels
with large delay spreads. In addition, no theoretical sparse approximation
guarantees are provided. 

In this paper, we develop a general framework\footnote{This framework generalizes our results in \cite{ourFWg} from linear
equalization (which follows as a special case of our framework presented
here by setting the TIR to be a single delayed impulse) to CSE. } for the design of sparse CSE and TIR FIR filters that transforms
the original problem into one of sparse approximation of a vector
using different dictionaries. The developed framework can then be
used to find the sparsifying dictionary that leads to the sparsest
FIR filter subject to an approximation constraint. Moreover, we investigate
the coherence of the sparsifying dictionaries that we propose as part
of our analysis and identify one dictionary that has the smallest
coherence. Then, we use simulations to validate that the dictionary
with the smallest coherence results in the sparsest FIR design. Finally,
the numerical results demonstrate the significance of our approach
compared to conventional sparse TIR designs, e.g., in \cite{sigTaps},
in terms of both performance and computational complexity.

\textbf{\textit{Notations}}: We use the following standard notation
in this paper: $\mbox{\ensuremath{\boldsymbol{I}}}_{N}$ denotes the
identity matrix of size $N$. Upper- and lower-case bold letters denote
matrices and vectors, respectively. Underlined upper-case bold letters,
e.g., $\boldsymbol{\underline{X}}$, denote frequency-domain vectors.
The notations $(.)^{-1},\,(.)^{*},\,(.)^{T}\mbox{ and }\,(.)^{H}$
denote the matrix inverse, the matrix (or element) complex conjugate,
the matrix transpose and the complex-conjugate transpose operations,
respectively. $E\left[.\right]$ denotes the expected value operator.
$\left\Vert .\right\Vert _{\ell}$ and $\left\Vert .\right\Vert _{F}$
denote the $\ell$-norm and Frobenius norm, respectively. $\otimes$
denotes the Kronecker product of matrices. The components of a vector
starting from $k_{1}$ and ending at $k_{2}$ are given as subscripts
to the vector separated by a colon, i.e., $\boldsymbol{x}_{k_{1}:k_{2}}.$

\section{System Model And Assumptions\label{sub:Signal-Model}}

\begin{figure}[t]
\centering\includegraphics[scale=0.6]{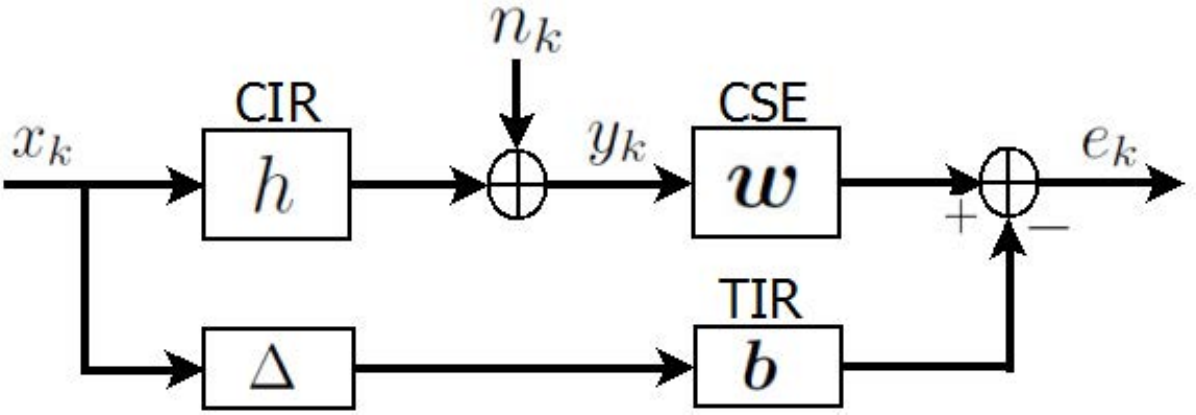}

\vspace{-0.5em}

\protect\caption{{\small{}A schematic of the system model. }\label{fig:Structure-of-the}}

\vspace{-.5em}
\end{figure}

A schematic of the system model studied in this paper is shown in
Figure \ref{fig:Structure-of-the}. We assume a linear, time-invariant,
dispersive and noisy communication channel. The standard complex-valued
equivalent baseband signal model is assumed. At time $k$, the received
sample $y_{k}$ is given by

\vspace{-1.0em}

\begin{equation}
y_{k}=\sum_{l=0}^{v}h_{l}\,x_{k-l}\,+n_{k},\label{eq:y_k}
\end{equation}
where $h_{l}$ is the CIR whose memory is $v$, $n_{k}$ is the additive
noise symbol and $x_{k-l}$ is the transmitted symbol at time ($k-l$).
We assume a symbol-spaced CSE but our proposed design framework can
be easily extended to the general fractionally-spaced case. Over a
block of $N_{f}$ output samples, the input-output relation in (\ref{eq:y_k})
can be written in a compact form as 

\vspace{-1.0em}

\begin{equation}
\boldsymbol{y}_{k:k-N_{f}+1}=\boldsymbol{H}\,\boldsymbol{x}_{k:k-N_{f}-v+1}+\boldsymbol{n}_{k:k-N_{f}+1}\,,\label{eq:y_Hx_n}
\end{equation}
where $\boldsymbol{y}_{k:k-N_{f}+1},\,\boldsymbol{x}_{k:k-N_{f}-v+1}$
and $\boldsymbol{n}_{k:k-N_{f}+1}$ are column vectors grouping the
received, transmitted and noise samples. Furthermore, $\boldsymbol{H}$
is an $N_{f}\times(N_{f} + v)$ Toeplitz matrix whose first row is
formed by $\{\boldsymbol{h}_{l}\}_{l=0}^{l=v}$ followed by zero entries.
It is useful, as will be shown in the sequel, to define the output
auto-correlation and the input-output cross-correlation matrices based
on the block of length $N_{f}$. Using (\ref{eq:y_Hx_n}), the input
correlation and the noise correlation matrices are, respectively,
defined by {\small{}$\boldsymbol{R}_{xx}\triangleq E\left[\boldsymbol{x}_{k:k-N_{f}-v+1}\boldsymbol{x}_{k:k-N_{f}-v+1}^{H}\right]\mbox{ and }\boldsymbol{R}_{nn}\triangleq E\left[\boldsymbol{n}_{k:k-N_{f}+1}\boldsymbol{n}_{k:k-N_{f}-1}^{H}\right]$}.
Both the input and noise processes are assumed to be white; hence,
their auto-correlation matrices are assumed to be (multiples of) the
identity matrix, i.e., $\boldsymbol{R}_{xx}=\boldsymbol{I}_{N_{f}+v}$
and $\boldsymbol{R}_{nn}=\frac{1}{SNR}\boldsymbol{I}_{N_{f}}$. Moreover,
the output-input cross-correlation and the output auto-correlation
matrices are, respectively, defined as

\vspace{-1.0em}

{\small{}
\begin{eqnarray}
\boldsymbol{R}_{yx} & \!\!\triangleq & \!\!E\left[\boldsymbol{y}_{k:k-N_{f}+1}\boldsymbol{x}_{k:k-N_{f}-v+1}^{H}\right]=\boldsymbol{H}\boldsymbol{R}_{xx}\,,\,\mbox{and}\\
\boldsymbol{R}_{yy} & \!\!\triangleq & \!\!E\left[\boldsymbol{y}_{k:k-N_{f}+1}\boldsymbol{y}_{k:k-N_{f}+1}^{H}\right]\!\!=\boldsymbol{H}\boldsymbol{R}_{xx}\boldsymbol{H}^{H}+\boldsymbol{R}_{nn}.\label{eq:R_yy_def}
\end{eqnarray}
}{\small \par}

\section{Sparse Channel Shortening and Equalization\label{sec:Sparse-FIR-Equalization}}

\subsection{Problem Formulation}

As shown in Figure \ref{fig:Structure-of-the}, in minimum-mean-square-error
(MMSE) shortening \cite{effcompRed}, the objective is to design the
CSE, $\boldsymbol{w}$, and the TIR, $\boldsymbol{b}$, filters such
that the overall impulse response which is the convolution of the
CIR and the CSE vectors best approximates, in the MSE sense, a TIR
with few taps. We denote the span (the filter length) of the CSE by
$N_{f}$ and the span of the TIR by $N_{b}$. The $k^{th}$ channel
shortening error sample $e_{k}$ is defined as follows (see Figure
\ref{fig:Structure-of-the}):

{\small{}
\begin{eqnarray}
e_{k} & = & \sum_{m=0}^{N_{f}-1}w_{n}^{*}\boldsymbol{y}_{k-m}-\sum_{n=0}^{N_{b}}b_{n}^{*}\boldsymbol{x}_{k-\Delta-n}\nonumber \\
 & = & \underbrace{\left[\begin{array}{cccc}
w_{0}^{*} & w_{1}^{*} & \ldots & w_{N_{f}-1}^{*}\end{array}\right]}_{\boldsymbol{w}^{H}}\boldsymbol{y}_{k:k+N_{f}+1}-\nonumber \\
 &  & \underbrace{\left[\begin{array}{cccccc}
\boldsymbol{0}_{1\times\Delta} & b_{0}^{*} & b_{1}^{*} & \ldots & b_{N_{b}}^{*} & \boldsymbol{0}_{1\times s}\end{array}\right]}_{\boldsymbol{b}^{H}}\boldsymbol{x}_{k:k-N_{f}-v+1}\nonumber \\
 & = & \boldsymbol{w}^{H}\boldsymbol{y}_{k:k+N_{f}+1}-\boldsymbol{b}^{H}\boldsymbol{x}_{k:k-N_{f}-v+1}\,,
\end{eqnarray}
}where $\Delta$ represents the decision delay and $s=N_{f}+v-\Delta-N_{b}-1$.
Hence, the MSE, denoted as $\xi\left(\boldsymbol{w,b}\right)$ is
given by

{\small{}
\begin{eqnarray}
\!\!\!\!\!\!\!\xi\left(\boldsymbol{w,b}\right) & \triangleq & E\left[\left|e_{k}^{2}\right|\right]\nonumber \\
\!\!\!\!\!\!\! & = & \!\!\boldsymbol{w}^{H}\boldsymbol{R}_{yy}\boldsymbol{w}-\boldsymbol{w}^{H}\boldsymbol{R}_{yx}\boldsymbol{b}-\boldsymbol{b}\boldsymbol{R}_{yx}^{H}\boldsymbol{w}^{H}+\boldsymbol{b}^{H}\boldsymbol{R}_{xx}\boldsymbol{b}.\label{eq:mse_1}
\end{eqnarray}
}Using the well-known orthogonality principle for MMSE, i.e., the
error sequence is uncorrelated with the observed data, $E\left[e_{k}\boldsymbol{y}_{k:k+N_{f}+1}\right]=0$,
we get

\begin{equation}
\boldsymbol{w}^{H}\boldsymbol{R}_{yy}=\boldsymbol{b}^{H}\boldsymbol{R}_{yx}.\label{eq:w_cse}
\end{equation}

Combining (\ref{eq:mse_1}) and (\ref{eq:w_cse}), we derive the MSE
as 

\begin{eqnarray}
\xi\left(\boldsymbol{b}\right) & = & \boldsymbol{b}^{H}\underbrace{\left(\boldsymbol{R}_{xx}-\boldsymbol{R}_{xy}\boldsymbol{R}_{yy}^{-1}\boldsymbol{R}_{yx}\right)}_{\boldsymbol{R}_{\delta}}\boldsymbol{b}.\label{eq:err_min_b}
\end{eqnarray}

Minimizing $\xi\left(\boldsymbol{b}\right)$ over $\boldsymbol{b}$,
gives the trivial solution $\boldsymbol{b}=0$. Thus, to preclude
this trivial case, we minimize $\xi\left(\boldsymbol{b}\right)$ subject
to the UTC\footnote{The UTC is a generalization of the monicity constraint which is, in
general, imposed on linear prediction filters \cite{Haykin_Adp_filter}
where it results in a white error sequence $e_{k}$ when the length
of the filters increases, i.e., $N_{b}$ converges to $v$ and $N_{f}$
converges to infinity.  } where one of the $\left(N_{b}+1\right)$ nonzero taps is set to unity.
The index of this unit tap $i$ $\left(0<i\leq N_{f}+v-1\right)$
is chosen such that the MSE $\xi\left(\boldsymbol{b}\right)$ is minimized.
To minimize the MSE subject to the UTC, $\boldsymbol{b}^{H}\boldsymbol{e}_{i}=1$
(where $\boldsymbol{e}_{i}$ is the $i^{th}$ unit vector), we express
$\boldsymbol{R}_{\delta}$ in (\ref{eq:err_min_b}) as $\boldsymbol{A}^{H}\boldsymbol{A}$
where $\boldsymbol{A}$ is the square root matrix of $\boldsymbol{R}_{\delta}$
in the spectral-norm sense, which results from Cholesky or eigen decomposition
\cite{matAnalysis}. Then, (\ref{eq:err_min_b}) can be written as 

\begin{eqnarray}
\xi\left(\boldsymbol{b},i\right) & = & \boldsymbol{b}^{H}\boldsymbol{A}^{H}\boldsymbol{A}\boldsymbol{b}=\left\Vert \boldsymbol{A}\boldsymbol{b}\right\Vert _{2}^{2}\nonumber \\
 & = & \left\Vert \boldsymbol{\widetilde{A}}\,\widetilde{\boldsymbol{b}}+\boldsymbol{a}_{i}\right\Vert _{2}^{2}\,,\label{eq:quad_mse_TIR_b}
\end{eqnarray}
where $\boldsymbol{a}_{i}$ is the $i^{th}$ column of $\boldsymbol{A}$,
$\boldsymbol{\widetilde{A}}$ is composed of all columns of $\boldsymbol{A}$
except $\boldsymbol{a}_{i}$, and $\widetilde{\boldsymbol{b}}$ is
formed by all elements of $\boldsymbol{b}$ except the $i^{th}$ entry
with unit value. Since only $\left(N_{b}+1\right)$ taps out of the
total $\left(N_{f}+v\right)$ TIR taps are nonzero, the locations
and weights of these taps need to be estimated such that $\xi\left(\boldsymbol{b}\right)$
is minimized. Towards this goal, we need first to optimize the location
of the unit tap index $i$. By formulating the Lagrangian function:
$\boldsymbol{\mathcal{L}}^{UTC}\left(\boldsymbol{b},\lambda\right)=\xi\left(\boldsymbol{b}\right)+\lambda\left(\boldsymbol{b}^{H}\boldsymbol{e}_{i}-1\right)$,
where $\lambda$ is the Lagrangian multiplier, and then performing
the minimization with respect to $\boldsymbol{b}$, we get \cite{effcompRed}

{\small{}
\begin{equation}
\boldsymbol{b}_{opt}=\frac{\boldsymbol{R}_{\delta}^{-1}e_{i_{opt}}}{\boldsymbol{R}_{\delta}^{-1}\left(i_{opt},i_{opt}\right)}\,\,,\,\mbox{and}\,\,\,\,\,\mbox{MMSE}=\frac{1}{\boldsymbol{R}_{\delta}^{-1}\left(i_{opt},i_{opt}\right)}\,.\label{eq:b_opt_mmse_utc}
\end{equation}
}Note that $i_{opt}$ is the index that achieves the MMSE and is given
by 

{\small{}
\begin{equation}
i_{opt}\triangleq\underset{0<i\leq N_{f}+v-1}{\mbox{arg}\mbox{max}}\boldsymbol{R}_{\delta}^{-1}\left(i,i\right).
\end{equation}
}{\small \par}

After computing $i_{opt}$, we formulate the following problem for
the design of a sparse TIR filter:

\vspace{-1.0em}

\begin{eqnarray}
\!\!\!\widetilde{\boldsymbol{b}}_{s} & \triangleq & \underset{\boldsymbol{b}\in\mathbb{C}^{N_{f}+v}}{\mbox{arg}\mbox{min}}\,\,\left\Vert \boldsymbol{b}\right\Vert _{0}\,\,\,\,\mbox{subject to}\,\,\,\,\,\xi\left(\boldsymbol{b},i_{opt}\right)\leq\delta_{cs}\,,\label{eq:opt_prob1}
\end{eqnarray}
where $\left\Vert \boldsymbol{b}\right\Vert _{0}$ is the number of
nonzero elements in its argument and the threshold $\delta_{cs}$
can be used as a design parameter to control the performance-complexity
tradeoff. Although one can attempt to use convex-optimization-based
approaches (after replacing $\left\Vert .\right\Vert _{0}$ with its
convex approximation $\left\Vert .\right\Vert _{1}$ in (\ref{eq:opt_prob1})
to reduce the search space \cite{justRelax06}) for the sparse approximation
vector, $\widetilde{\boldsymbol{b}}_{s}$, a number of greedy algorithms
can be used to solve this problem efficiently with low complexity,
especially in the situations where a specific number of $N_{b}$ taps
is desired.

Once $\widetilde{\boldsymbol{b}}_{s}$ is calculated, we insert the
unit tap in the $i^{th}$ location to construct the sparse TIR, $\boldsymbol{b}_{s}$.
Then, the optimum CSE taps (in the MMSE sense) are determined from
(\ref{eq:w_cse}) to be 

\vspace{-1.0em}

\begin{equation}
\boldsymbol{w}_{opt}=\boldsymbol{R}_{yy}^{-1}\overbrace{\boldsymbol{R}_{yx}\boldsymbol{b}_{s}}^{\boldsymbol{t}}.\label{eq:opt_w_cse}
\end{equation}

Notice that the MMSE CSE, $\boldsymbol{w}_{opt}$, is not sparse and,
hence, its design complexity still increases proportional to $(N_{f})^{2}$,
which can be computationally expensive \cite{DCProakis}. In contrast,
any choice for $\boldsymbol{w}$ other than $\boldsymbol{w}_{opt}$
leads to performance loss, especially for channels with large delay
spreads. Therefore, we also propose a sparse implementation for the
FIR CSE, $\boldsymbol{w}$, as follows. After computing the TIR coefficients,
$\boldsymbol{b}_{s}$, the MSE will be a function only of $\boldsymbol{w}$
and can be expressed as 

\vspace{-0.5em}

{\small{}
\begin{eqnarray}
\xi(\boldsymbol{w}) & = & \boldsymbol{w}^{H}\boldsymbol{R}_{yy}\boldsymbol{w}-\boldsymbol{w}^{H}\boldsymbol{R}_{yx}\boldsymbol{b}_{s}-\boldsymbol{b}_{s}\boldsymbol{R}_{yx}^{H}\boldsymbol{w}^{H}+\boldsymbol{b}_{s}^{H}\boldsymbol{R}_{xx}\boldsymbol{b}_{s}\nonumber \\
\nonumber \\
\!\! & = & \!\underbrace{\boldsymbol{b}_{s}^{H}\boldsymbol{R}_{\delta}\boldsymbol{b}_{s}}_{\xi_{min}}+\underbrace{(\boldsymbol{w}-\boldsymbol{R}_{yy}^{-1}\boldsymbol{t})^{H}\boldsymbol{R}_{yy}(\boldsymbol{w}-\boldsymbol{R}_{yy}^{-1}\boldsymbol{t})}_{\xi_{ex}\left(\boldsymbol{w}\right)}\,.\label{eq:error_2}
\end{eqnarray}
}The MSE $\xi\left(\boldsymbol{w}\right)$ is minimized by minimizing
the term $\xi_{ex}(\boldsymbol{w})$ since $\xi_{min}$ does not depend
on $\boldsymbol{w}$. Hence, the optimum choice for $\boldsymbol{w}$,
in the MMSE sense, is the one given in (\ref{eq:opt_w_cse}). However,
its computation and implementation complexity are high. Alternatively,
to design a sprase CSE, we propose to use the excess MSE term, $\xi_{ex}\left(\boldsymbol{w}\right)$,
as a design constraint to once again achieve a desirable performance-complexity
tradeoff. In particular, we formulate the following problem for the
design of sparse FIR CSE, $\boldsymbol{w}$:

\vspace{-0.5em}

\begin{eqnarray}
\widehat{\boldsymbol{w}}_{s} & \triangleq & \underset{\boldsymbol{w}\in\mathbb{C}^{N_{f}}}{\mbox{arg}\mbox{min}}\,\,\left\Vert \boldsymbol{w}\right\Vert _{0}\,\,\,\,\mbox{subject to}\,\,\,\,\,\xi_{ex}\left(\boldsymbol{w}\right)\leq\delta_{eq}\,,\label{eq:opt_w_cse_1}
\end{eqnarray}
where $\delta_{eq}$ is a parameter selected by the designer to control
the performance loss from the non-sparse highly-complex conventional
CSE design based on the MMSE criterion. To solve (\ref{eq:opt_w_cse_1}),
we can use either convex optimization or greedy algorithms with the
latter being more desirable due to their low complexity. We are now
ready to discuss a general framework for designing sparse CSE and
TIR FIR filters such that the performance loss does not exceed an
acceptable predefined limit.

\vspace{-0.5em}

\subsection{Proposed Sparse Approximation Framework\label{sub:Proposed-sparse-approximation}}

Unlike earlier works, including the one by one of the co-authors \cite{newDFW},
we provide a general framework for designing both sparse CSE and TIR
FIR filters that can be considered as the problem of sparse approximation
using different dictionaries. Mathematically, this framework poses
the design problem as follows:

\vspace{-1.5em}

\begin{equation}
\widehat{\boldsymbol{z}}_{s}\triangleq\underset{\boldsymbol{z}}{\mbox{\mbox{arg}\mbox{min}}}\,\left\Vert \boldsymbol{z}\right\Vert _{0}\,\,\,\mbox{subject to}\,\,\,\left\Vert \boldsymbol{K}\left(\boldsymbol{\varPhi}\boldsymbol{z}-\boldsymbol{d}\right)\right\Vert _{2}^{2}\leq\epsilon\,,\label{eq:propFW}
\end{equation}
where $\boldsymbol{\varPhi}$ is the dictionary that will be used
to sparsely approximate $\boldsymbol{d}$, while $\boldsymbol{K}$
is a known matrix and $\boldsymbol{d}$ is a known data vector, both
of which change depending upon the sparsifying dictionary $\boldsymbol{\varPhi}$.
Notice that $\widehat{\boldsymbol{z}}_{s}\in\left\{ \widehat{\boldsymbol{w}}_{s},\,\widetilde{\boldsymbol{b}}_{s}\right\} $
and $\epsilon\in\left\{ \delta_{cs},\,\delta_{eq}\right\} $. Notice
that also by completing the square in (\ref{eq:opt_w_cse_1}) the
problem reduces to the one shown in (\ref{eq:propFW}). Hence, one
can use any factorization for $\boldsymbol{R}_{\delta}$ in (\ref{eq:err_min_b})
or $\boldsymbol{R}_{yy}$ in (\ref{eq:error_2}) to formulate a sparse
approximation problem. Using the Cholesky or eigen decomposition for
$\boldsymbol{R}_{\delta}$ or $\boldsymbol{R}_{yy}$, we will have
different choices for $\boldsymbol{K}$, $\boldsymbol{\varPhi}$ and
$\boldsymbol{d}$. For instance, by defining the Cholesky factorization
\cite{matAnalysis} of $\boldsymbol{R}_{\delta}$ as $\boldsymbol{R}_{\delta}\triangleq\boldsymbol{L}_{\delta}\boldsymbol{L}_{\delta}^{H}$,
or in the equivalent form $\boldsymbol{R}_{\delta}\triangleq\boldsymbol{P}_{\delta}\boldsymbol{\Sigma}_{\delta}\boldsymbol{P}_{\delta}^{H}=\boldsymbol{\varOmega}_{\delta}\boldsymbol{\varOmega}_{\delta}^{H}$
(where $\boldsymbol{L}_{\delta}$ is a lower-triangular matrix, $\boldsymbol{P}_{\delta}$
is a lower-unit-triangular (unitriangular) matrix and $\boldsymbol{\Sigma}_{\delta}$
is a diagonal matrix), the problem in (\ref{eq:propFW}) can, respectively,
take one of the forms shown below:

{\small{}
\begin{eqnarray}
 & \underset{\boldsymbol{b}\in\mathbb{C}^{N_{f}+v}}{\mbox{min}}\,\,\left\Vert \boldsymbol{b}\right\Vert _{0}\mbox{\,\,\,\,\mbox{s.t. }\,\,\,\,\ensuremath{\left\Vert \left(\widetilde{\boldsymbol{L}}_{\delta}^{H}\,\widetilde{\boldsymbol{b}}+\boldsymbol{l}_{i}\right)\right\Vert _{2}^{2}\leq\delta_{cs}\,},}\,\mbox{and}\label{eq:ex_R_delta_1}\\
 & \underset{\boldsymbol{b}\in\mathbb{C}^{N_{f}+v}}{\mbox{min}}\,\,\left\Vert \boldsymbol{b}\right\Vert _{0}\mbox{\,\,\,\,\mbox{s.t. }\,\,\,\,\ensuremath{\left\Vert \left(\widetilde{\boldsymbol{\varOmega}}_{\delta}^{H}\,\widetilde{\boldsymbol{b}}+\boldsymbol{p}_{i}\right)\right\Vert _{2}^{2}\leq\delta_{cs}\,}.}\label{eq:ex_R_delta_2}
\end{eqnarray}
}{\small \par}

Recall that $\widetilde{\boldsymbol{\varOmega}}_{\delta}^{H}$ is
formed by all columns of $\boldsymbol{\varOmega}_{\delta}^{H}$ except
the $i^{th}$ column, $\boldsymbol{p}_{i}$ is the $i^{th}$ column
of $\boldsymbol{\varOmega}_{\delta}^{H}$, and $\widetilde{\boldsymbol{b}}$
is formed by all entries of $\boldsymbol{b}$ except the $i^{th}$
unity entry. Similarly, by writing the Cholesky factorization of $\boldsymbol{R}_{yy}$
as $\boldsymbol{R}_{yy}\triangleq\boldsymbol{L}_{y}\boldsymbol{L}_{y}^{H}$
or the eigen decomposition of $\boldsymbol{R}_{yy}$ as $\boldsymbol{R}_{yy}\triangleq\boldsymbol{U}_{y}\boldsymbol{D}_{y}\boldsymbol{U}_{y}^{H}$,
we can formulate the problem in (\ref{eq:propFW}) as follows: 

{\small{}
\begin{eqnarray}
 & \!\!\!\!\underset{\boldsymbol{w}\in\mathbb{C}^{N_{f}}}{\mbox{min}}\left\Vert \boldsymbol{w}\right\Vert _{0}\mbox{\,\,\,\,\mbox{s.t. }\,\,\,\,\ensuremath{\left\Vert \left(\boldsymbol{L}_{y}^{H}\boldsymbol{w}-\boldsymbol{L}_{y}^{-1}\boldsymbol{t}\right)\right\Vert _{2}^{2}\leq\delta_{eq}\,},}\label{eq:L_y_h}\\
 & \!\!\!\!\!\!\!\!\!\!\!\,\underset{\boldsymbol{w}\in\mathbb{C}^{N_{f}}}{\mbox{min}}\left\Vert \boldsymbol{w}\right\Vert _{0}\mbox{\,\,\mbox{s.t. }\,\,\ensuremath{\left\Vert \left(\boldsymbol{D}_{y}^{\frac{1}{2}}\boldsymbol{U}_{y}^{H}\boldsymbol{w}-\boldsymbol{D}_{y}^{-\frac{1}{2}}\boldsymbol{U}_{y}^{H}\boldsymbol{t}\right)\right\Vert _{2}^{2}\leq\delta_{eq},\,\mbox{and}}}\label{eq:U_telda_y}\\
 & \!\!\!\!\!\!\underset{\boldsymbol{w}\in\mathbb{C}^{N_{f}}}{\mbox{min}}\left\Vert \boldsymbol{w}\right\Vert _{0}\mbox{\,\,\,\,\mbox{s.t. }\,\,\,\,\ensuremath{\ensuremath{\left\Vert \boldsymbol{L}_{y}^{-1}\left(\boldsymbol{R}_{yy}\boldsymbol{w}-\boldsymbol{t}\right)\right\Vert _{2}^{2}}\leq\delta_{eq}}\,}.\label{eq:R_yy_min_prob_cse}
\end{eqnarray}
}{\small \par}

Note that the sparsifying dictionaries in (\ref{eq:L_y_h}), (\ref{eq:U_telda_y})
and (\ref{eq:R_yy_min_prob_cse}) are $\boldsymbol{L}_{y}^{H}$, $\boldsymbol{D}_{y}^{\frac{1}{2}}\boldsymbol{U}_{y}^{H}$
and $\boldsymbol{R}_{yy}$, respectively. Furthermore, the matrix
$\boldsymbol{K}$ is an identity matrix in all cases except in (\ref{eq:R_yy_min_prob_cse}),
where it is equal to $\boldsymbol{L}_{y}^{-1}$. 

We conclude this section by pointing out that several other sparsifying
dictionaries can be used to sparsely design the CSE and TIR FIR filters.
Due to lack of space, we presented above some of those possible choices
and the other choices can be derived by applying suitable transformations
to (\ref{eq:err_min_b}) and (\ref{eq:error_2}).

\subsection{Reduced-Complexity Design }

In this section, we propose reduced-complexity designs for the CSE
and TIR FIR filters. The proposed designs in Section \ref{sub:Proposed-sparse-approximation}
involve Cholesky factorization and/or eigen decomposition, whose computational
costs could be large for channels with large delay spreads. For a
Toeplitz matrix, the most efficient algorithms for Cholesky factorization
are Levinson or Schur algorithms \cite{statDSP}, which involve $\mathcal{O}(M^{2})$
computations, where $M$ is the matrix dimension. In contrast, since
a circulant matrix is asymptotically equivalent to a Toeplitz matrix,
for reasonably large dimension \cite{toep2circApp2003}, the eigen
decomposition of a circulant matrix can be computed efficiently using
the fast Fourier transform (FFT) and its inverse with only $\mathcal{O}\left(M\,\mbox{log}(M)\right)$
operations. By using this asymptotic equivalence between Toeplitz
and circulant matrices, all computations needed for $\boldsymbol{R}_{yy}$
and $\boldsymbol{R}_{\delta}$ factorization can be done efficiently
with the FFT and inverse FFT. In addition, direct matrix inversion
can be avoided.

It is well known that a circulant matrix, $\boldsymbol{C}$, has the
discrete Fourier transform (DFT) basis vectors as its eigenvectors
and the DFT of its first column as its eigenvalues. An $M\times M$
circulant matrix $\boldsymbol{C}$ can be decomposed as $\boldsymbol{C}$
=$\frac{1}{M}\boldsymbol{F}_{M}^{H}\boldsymbol{\varLambda}_{\boldsymbol{c}}\boldsymbol{F}_{M}${\small{},
}where $\boldsymbol{F}_{M}$ is the discrete Fourier transform (DFT)
matrix with $f_{k,l}=e^{-j2\pi kl/M}$, $0\leq k,\,l\leq M-1$, and
$\boldsymbol{\varLambda}_{\boldsymbol{c}}$ is an $M$ by $M$ diagonal
matrix whose diagonal elements are the $M$-point DFT of $\boldsymbol{c}=\left\{ c\right\} _{i=0}^{i=M-1}$,
the first column of the circulant matrix. We denote by $\overline{\boldsymbol{R}}_{yy},\,\overline{\boldsymbol{R}}_{yx}$
and $\overline{\boldsymbol{R}}_{\delta}$ the circulant approximation
to the matrices $\boldsymbol{R}_{yy},\,\boldsymbol{R}_{yx}\mbox{ and }\boldsymbol{R}_{\delta}$,
respectively. In addition, we denote the noiseless channel output
vector as $\widetilde{\boldsymbol{y}},$ i.e., $\widetilde{\boldsymbol{y}}=\boldsymbol{Hx}$.
Hence, the autocorrelation matrix $\boldsymbol{\overline{R}}_{yy}$
is computed as 

{\small{}
\begin{equation}
\boldsymbol{\overline{R}}_{yy}=\underbrace{E\left[\widetilde{\boldsymbol{y}}_{k}\widetilde{\boldsymbol{y}}_{k}\right]}_{\overline{\boldsymbol{R}}_{\widetilde{y}\widetilde{y}}}+\underbrace{\frac{1}{SNR}}_{\sigma_{n}^{2}}\boldsymbol{I}_{N_{f}}.
\end{equation}
}{\small \par}

To approximate $\boldsymbol{R}_{yy}$ as a circulant matrix, we assume
that $\left\{ \widetilde{\boldsymbol{y}}_{k}\right\} $ is cyclic.
Hence, $E\left[\widetilde{\boldsymbol{y}}_{k}\widetilde{\boldsymbol{y}}_{k}\right]$
can be approximated as a time-averaged autocorrelation function as
follows:

\vspace{-1.5em}

{\small{}
\begin{eqnarray}
\overline{\boldsymbol{R}}_{\widetilde{y}\widetilde{y}} & = & \frac{1}{N_{f}}\sum_{k=0}^{N_{f}-1}\widetilde{\boldsymbol{y}}_{k}\widetilde{\boldsymbol{y}}_{k}^{H}=\frac{1}{N_{f}}\boldsymbol{C}_{y}\boldsymbol{C}_{y}^{H}\nonumber \\
 & = & \frac{1}{N_{f}}\left(\frac{1}{N_{f}}\boldsymbol{F}_{N_{f}}^{H}\boldsymbol{\varLambda}_{\underline{\widetilde{\boldsymbol{Y}}}}\boldsymbol{F}_{N_{f}}\right)\left(\frac{1}{N_{f}}\boldsymbol{F}_{N_{f}}^{H}\boldsymbol{\varLambda}_{\underline{\widetilde{\boldsymbol{Y}}}}^{H}\boldsymbol{F}_{N_{f}}\right)\nonumber \\
 & = & \frac{1}{N_{f}^{2}}\boldsymbol{F}_{N_{f}}^{H}\boldsymbol{\varLambda}_{\underline{\widetilde{\boldsymbol{Y}}}}\varLambda_{\underline{\widetilde{\boldsymbol{Y}}}}^{H}\boldsymbol{F}_{N_{f}}\nonumber \\
 & = & \frac{1}{N_{f}^{2}}\left(\boldsymbol{F}_{N_{f}}^{H}\boldsymbol{\varLambda}_{\left|\underline{\widetilde{\boldsymbol{Y}}}\right|^{2}}\boldsymbol{F}_{N_{f}}\right)\,,\label{eq:Ryy_circ}
\end{eqnarray}
}where $\underline{\widetilde{\boldsymbol{Y}}}$ is the $N_{f}$-point
DFT of $\boldsymbol{c}_{1}=\left[\begin{array}{cccc}
\widetilde{y}_{N_{f}-1} & \widetilde{y}_{N_{f}-2} & \ldots & \widetilde{y}_{0}\end{array}\right]$, $\boldsymbol{F}_{N_{f}}$ is an $N_{f}\times N_{f}$ DFT matrix.
$\left|.\right|$ denotes element-wise norm square, and $\boldsymbol{C}_{y}=\mbox{\textbf{circ}}\left(\boldsymbol{c}_{1}\right)$
where \textbf{circ} denotes a circulant matrix whose first column
is $\boldsymbol{c}_{1}${\footnotesize{}. }Note that $\boldsymbol{F}_{N_{f}}^{H}\boldsymbol{F}_{N_{f}}=\boldsymbol{F}_{N_{f}}\boldsymbol{F}_{N_{f}}^{H}=N_{f}\boldsymbol{I}_{N_{f}}${\footnotesize{}.}{\footnotesize \par}

Without loss of generality, we can write the noiseless channel output
sequence $\widetilde{\boldsymbol{y}}_{k}$ in the discrete frequency
domain as a column vector form as

\vspace{-1.0em}

{\small{}
\begin{eqnarray}
\widetilde{\boldsymbol{\underline{Y}}} & = & \boldsymbol{\underline{H}}^{H}\odot\boldsymbol{\underline{P}}_{\Delta}\odot\boldsymbol{\underline{X}}\,,
\end{eqnarray}
}where $\boldsymbol{\underline{H}}$ is the $N_{f}$-point DFT of
the CIR $\boldsymbol{h}$, $\boldsymbol{\underline{P}}_{\Delta}=\left[\begin{array}{cccc}
1 & e^{-j2\pi\Delta/N_{f}} & \ldots & e^{-j2\pi\left(N_{f}-1\right)\Delta/N_{f}}\end{array}\right]^{T}$, and $\odot$ denotes element-wise multiplication. Note that $\left|\widetilde{\boldsymbol{\underline{Y}}}\right|^{2}=N_{f}\left|\boldsymbol{\underline{H}}\right|^{2}$.
Then,

\vspace{-1.5em}

{\small{}
\begin{eqnarray}
\boldsymbol{\overline{R}}_{yy} & = & \overline{\boldsymbol{R}}_{\widetilde{y}\widetilde{y}}+\sigma_{n}^{2}\boldsymbol{I}_{N_{f}}\nonumber \\
 & = & \frac{1}{N_{f}^{2}}\boldsymbol{F}_{N_{f}}^{H}\left(\boldsymbol{\varLambda}_{N_{f}\left|\boldsymbol{\underline{H}}\right|^{2}+N_{f}\sigma_{n}^{2}\boldsymbol{1}_{N_{f}}}\right)\boldsymbol{F}_{N_{f}}\nonumber \\
 & = & \boldsymbol{F}_{N_{f}}^{H}\left(\frac{1}{N_{f}}\boldsymbol{\varLambda}_{\underbrace{\left|\underline{\boldsymbol{H}}\right|^{2}+\sigma_{n}^{2}\boldsymbol{1}_{N_{f}}}_{\varrho}}\right)\boldsymbol{F}_{N_{f}}\nonumber \\
 & = & \boldsymbol{F}_{N_{f}}^{H}\left(\frac{1}{N_{f}}\boldsymbol{\varLambda}_{\varrho}\right)\boldsymbol{F}_{N_{f}}=\boldsymbol{Q}\boldsymbol{Q}^{H}.\label{eq:R_yy_circ}
\end{eqnarray}
}{\small \par}

Using the matrix inversion lemma \cite{matAnalysis}, the inverse
of $\boldsymbol{\overline{R}}_{yy}$ is then

\vspace{-1.5em}

{\small{}
\begin{eqnarray}
\boldsymbol{\overline{R}}_{yy}^{-1} & \!\!\!\!= & \!\!\!\left\{ \boldsymbol{F}_{N_{f}}^{H}\left(\frac{1}{N_{f}}\boldsymbol{\varLambda}_{\varrho}\right)\boldsymbol{F}_{N_{f}}\right\} ^{-1}\nonumber \\
 & \!\!\!\!\!= & \!\!\!\!\boldsymbol{F}_{N_{f}}^{H}\!\left[\!\frac{1}{N_{f}\sigma_{n}^{2}}\!\left(\!\boldsymbol{I}_{N_{f}}-\frac{1}{N_{f}\sigma_{n}^{2}}\boldsymbol{\varLambda}_{\underline{\widetilde{\boldsymbol{Y}}}}\!\left(\!\boldsymbol{\varLambda}_{\underline{\widetilde{\boldsymbol{Y}}}}^{H}\varLambda_{\underline{\widetilde{\boldsymbol{Y}}}}+\boldsymbol{I}_{N_{f}}\!\right)^{-1}\!\!\boldsymbol{F}_{N_{f}}\!\right)\!\right]\nonumber \\
\nonumber \\
 & \!\!\!\!= & \!\!\!\frac{1}{N_{f}\sigma_{n}^{2}}\boldsymbol{F}_{N_{f}}^{H}\left(\boldsymbol{I}_{N_{f}}-\frac{1}{N_{f}}\boldsymbol{\varLambda}_{\underline{\widetilde{\boldsymbol{Y}}}}\boldsymbol{\varLambda}_{\varrho}^{-1}\boldsymbol{\varLambda}_{\underline{\widetilde{\boldsymbol{Y}}}}^{H}\right)\boldsymbol{F}_{N_{f}}.\label{eq:R_yy_inv_circ}
\end{eqnarray}
}{\small \par}

Similarly, $\boldsymbol{R}_{\delta}$ can be expressed as 

\vspace{-1.0em}

{\small{}
\begin{eqnarray}
\overline{\boldsymbol{R}}_{\delta} & = & \boldsymbol{R}_{xx}-\boldsymbol{\overline{R}}_{yx}^{H}\boldsymbol{\overline{R}}_{yy}^{-1}\boldsymbol{\overline{R}}_{yx}\nonumber \\
 & = & \boldsymbol{R}_{xx}-\boldsymbol{\overline{R}}_{yx}^{H}\boldsymbol{\overline{R}}_{yy}^{-1}\left\{ \frac{1}{N^{2}}\widetilde{\boldsymbol{F}}_{N_{f}}^{H}\left(\boldsymbol{\varLambda}_{\underline{\widetilde{\boldsymbol{Y}}}}\boldsymbol{\varLambda}_{\boldsymbol{\underline{X}}}^{H}\right)\boldsymbol{F}_{N}\right\} \nonumber \\
 & = & \boldsymbol{I}_{N}-\boldsymbol{\overline{R}}_{yx}^{H}\times\nonumber \\
 &  & \left\{ \frac{1}{N^{2}\sigma_{n}^{2}}\widetilde{\boldsymbol{F}}_{N_{f}}^{H}\left(\boldsymbol{\varLambda}_{\underline{\boldsymbol{Y}}}\boldsymbol{\varLambda}_{\boldsymbol{\underline{X}}}^{H}-\boldsymbol{\varLambda}_{\boldsymbol{\underline{Y}}}\boldsymbol{\varLambda}_{\overline{\theta}\varoslash\theta}\boldsymbol{\varLambda}_{\boldsymbol{\underline{X}}}^{H}\right)\boldsymbol{F}_{N}\right\} \nonumber \\
 & = & \boldsymbol{I}_{N}-\boldsymbol{\overline{R}}_{xy}\left\{ \frac{1}{N\sigma_{n}^{2}}\boldsymbol{F}_{N}^{H}\left(\boldsymbol{\varLambda}_{\underline{\widetilde{\boldsymbol{Y}}}}\boldsymbol{\varLambda}_{1\varoslash\theta}\boldsymbol{\varLambda}_{\underline{\widetilde{\boldsymbol{Y}}}}^{H}\right)\boldsymbol{F}_{N}\right\} \nonumber \\
 & = & \boldsymbol{I}_{N}-\left\{ \frac{1}{N^{2}}\boldsymbol{F}_{N}^{H}\left(\boldsymbol{\varLambda}_{\boldsymbol{\underline{X}}}\boldsymbol{\varLambda}_{\underline{\widetilde{\boldsymbol{Y}}}}^{H}\boldsymbol{\varLambda}_{\underline{\widetilde{\boldsymbol{Y}}}}\boldsymbol{\varLambda}_{1\varoslash\theta}\boldsymbol{\varLambda}_{\boldsymbol{\underline{X}}}^{H}\right)\boldsymbol{F}_{N}\right\} \nonumber \\
 & = & \frac{1}{N^{2}}\boldsymbol{F}_{N}^{H}\left(N\,\boldsymbol{I}_{N}-\boldsymbol{\varLambda}_{\boldsymbol{\underline{X}}}\boldsymbol{\varLambda}_{\overline{\theta}\varoslash\theta}\boldsymbol{\varLambda}_{\boldsymbol{\underline{X}}}^{H}\right)\boldsymbol{F}_{N}\nonumber \\
 & = & \frac{1}{N}\boldsymbol{F}_{N}^{H}\left(\boldsymbol{I}_{N}-\boldsymbol{\varLambda}_{\overline{\theta}\varoslash\theta}\right)\boldsymbol{F}_{N}\nonumber \\
 & = & \boldsymbol{\varGamma}\boldsymbol{\varGamma}^{H},\label{eq:R_delta_circ}
\end{eqnarray}
}where $\varoslash$ denotes element-wise division, $N=N_{f}+v$,
$\widetilde{\boldsymbol{F}}_{N_{f}}^{H}$ is an $N\times N_{f}$ DFT
matrix, $\theta=\overline{\theta}+N\sigma_{n}^{2}\boldsymbol{1}_{N}$,
and $\overline{\theta}=\left|\widetilde{\boldsymbol{\underline{Y}}}\right|^{2}$.
Substituting (\ref{eq:R_yy_circ}) and (\ref{eq:R_delta_circ}) into
(\ref{eq:error_2}) and (\ref{eq:err_min_b}), respectively, we can
design the CSE and TIR FIR filters in a reduced-complexity manner
where neither a Choleskey nor an eigen factorization is needed. 

We have now shown that the problem of designing sparse CSE and TIR
FIR filters can be cast into one of sparse approximation of a vector
by a fixed dictionary. Furthermore, this dictionary can be obtained
in an efficient manner using only the FFT and its inverse. The general
form of this problem is given by (\ref{eq:propFW}). To solve this
problem, we use the well-known Orthogonal Matching Pursuit (OMP) greedy
algorithm \cite{omp07} that estimates $\widehat{\boldsymbol{z}}_{s}$
by iteratively selecting a set $S$ of the sparsifying dictionary
columns (i.e., atoms $\boldsymbol{\phi}_{i}$'s) of $\boldsymbol{\varPhi}$
that are most correlated with the data vector $\boldsymbol{d}$ and
then solving a restricted least-squares problem using the selected
atoms. The OMP stopping criterion ($\rho$) is traditionally either
a predefined sparsity level (number of nonzero entries) of $\boldsymbol{z_{s}}$
or an upper-bound on the norm of the residual error. In our problem,
$\rho$ in the latter case is changed from an upper-bound on the residual
error norm to an upper-bound on the norm of the Projected Residual
Error (PRE), i.e., ``$\boldsymbol{K}\times\mbox{Residual Error}$''.
The computations involved in the OMP algorithm are well documented
in the sparse approximation literature (e.g., \cite{omp07}) and are
omitted here due to page limitations. 

Note that unlike conventional compressive sensing techniques \cite{CS},
where the measurement matrix is a fat matrix, the sparsifying dictionary
in our framework is either a tall matrix (fewer columns than rows)
with full column rank as in (\ref{eq:ex_R_delta_1}) and (\ref{eq:ex_R_delta_2})
or a square one with full rank as in (\ref{eq:L_y_h})--(\ref{eq:R_yy_min_prob_cse}).
However, OMP and similar methods can still be used if $\boldsymbol{R}_{yy}$
and $\boldsymbol{R}_{\delta}$ can be decomposed into $\boldsymbol{\Psi}\boldsymbol{\Psi}^{H}$
and the data vector $\boldsymbol{d}$ is compressible \cite{sparsefeng2012,sparseFilterDesign13}. 

Our next challenge is to determine the best sparsifying dictionary
for use in our framework. We know from the sparse approximation literature
that the sparsity of the OMP solution tends to be inversely proportional
to the worst-case coherence $\mu\left(\boldsymbol{\varPhi}\right)$,
{\small{}$\mu\left(\boldsymbol{\varPhi}\right)\triangleq\underset{i\neq j}{\mbox{max}}\frac{\left|\left\langle \phi_{i},\,\phi_{j}\right\rangle \right|\,}{\left\Vert \phi_{i}\right\Vert _{2}\left\Vert \phi_{j}\right\Vert _{2}}$}
\cite{finiteSparseFilter013,greedIsGood03}. Notice that $\mu\left(\boldsymbol{\varPhi}\right)\in\left[0,1\right]$.
Next, we investigate the coherence of the dictionaries involved in
our analysis.

\subsection{Worst-Case Coherence Analysis\label{sub:Preliminary-Analysis} }

We perform a coherence metric analysis to gain some insight into the
performance of the proposed sparsifying dictionaries and the behavior
of the resulting sparse CSE and TIR FIR filters. First and foremost,
we are concerned with analyzing $\mu\left(\boldsymbol{\varPhi}\right)$
to ensure that it does not approach $1$ for any of the proposed sparsifying
dictionaries. In addition, we are interested in identifying which
$\boldsymbol{\varPhi}$ has the smallest coherence and, hence, gives
the sparsest design. We have two kinds of sparsifying dictionaries.
The first kind is the dictionaries resulting from $\boldsymbol{R}_{\delta}$
factorization while the second kind is either $\boldsymbol{R}_{yy}$
itself or any of its factors. We proceed as follows to characterize
upper-bounds on each kind of dictionary. We estimate upper bounds
on the worst-case coherence of both $\boldsymbol{R}_{\delta}$ and
$\boldsymbol{R}_{yy}$ separately and evaluate their closeness to
$1$. Then, through simulation, we demonstrate that the coherence
of the $\boldsymbol{R}_{\delta}$ and $\boldsymbol{R}_{yy}$ factors
will be less than that of $\mu(\boldsymbol{R}_{\delta})$ and $\mu(\boldsymbol{R}_{yy})$,
respectively. It is important to note here that the other dictionaries,
which result from decomposing $\boldsymbol{R}_{\delta}$ and $\boldsymbol{R}_{yy}$,
can be considered as square roots of them in the spectral-norm sense.
For example, $\left\Vert \boldsymbol{R}_{yy}\vphantom{\boldsymbol{L}^{H}}\right\Vert _{2}=\left\Vert \boldsymbol{L}_{y}\boldsymbol{L}_{y}^{H}\right\Vert _{2}\leq\left\Vert \boldsymbol{L}_{y}^{H}\boldsymbol{\vphantom{\boldsymbol{L}^{H}}}\right\Vert _{2}^{2}$
and $\left\Vert \boldsymbol{R}_{\delta}\vphantom{\boldsymbol{\boldsymbol{U}}_{\delta}^{H}}\right\Vert _{2}=\left\Vert \boldsymbol{\boldsymbol{U}}_{\delta}\boldsymbol{D}_{\delta}\boldsymbol{\boldsymbol{U}}_{\delta}^{H}\right\Vert _{2}\leq\left\Vert \boldsymbol{D}_{\delta}^{1/2}\boldsymbol{\boldsymbol{U}}_{\delta}^{H}\right\Vert _{2}^{2}$. 

Even though the matrix $\boldsymbol{R}_{\delta}$ is comprised of
Toeplitz matrices, it is generally non-Toeplitz. We use its factors
in our framework to compute $\widetilde{\boldsymbol{b}}$ that best
approximates $\boldsymbol{\widetilde{A}}\,\widetilde{\boldsymbol{b}}$
to $\boldsymbol{a}_{i}$ as shown in (\ref{eq:quad_mse_TIR_b}). The
formula of $\boldsymbol{R}_{\delta}$ in (\ref{eq:err_min_b}) can
be written in terms of the SNR and CIR coefficients as $\boldsymbol{R}_{\delta}=\left[\boldsymbol{R}_{xx}^{-1}+\boldsymbol{H}^{H}\boldsymbol{R}_{nn}^{-1}\boldsymbol{H}\right]^{-1}=\left[\boldsymbol{I}+\frac{1}{\sigma_{n}^{2}}\boldsymbol{H}^{H}\boldsymbol{H}\right]^{-1}$.
Hence, at low SNR, the noise effect dominates, i.e., $\boldsymbol{R}_{\delta}\approx\boldsymbol{I}$,
and, consequently, $\mu\left(\boldsymbol{R}_{\delta}\right)\rightarrow0$.
As the SNR increases, the noise effect decreases and the effect of
the channel taps starts to appear which makes $\mu\left(\boldsymbol{R}_{\delta}\right)$
converge to a constant. Through simulation, we show that this constant
does not approach $1$ and, accordingly, the other dictionaries generated
from $\boldsymbol{R}_{\delta}$ have a coherence that is also less
than that of $\mu\left(\boldsymbol{R}_{\delta}\right)$.

On the other hand, $\boldsymbol{R}_{yy}$ has a well-structured (Hermitian
Toeplitz) closed-form in terms of the CIR coefficients, filter time
span $N_{f}$ and SNR, i.e., $\boldsymbol{R}_{yy}=\boldsymbol{H}\boldsymbol{H}^{H}+\mbox{\ensuremath{\frac{\mbox{1}}{SNR}}}\boldsymbol{I}$.
It can be expressed in matrix form as 

\vspace{-1.5em}

{\small{}
\begin{equation}
\boldsymbol{R}_{yy}=\mbox{Toeplitz}\overbrace{\left(\left[\begin{array}{ccccccc}
r_{0} & r_{1} & \ldots & r_{v} & 0 & \ldots & 0\end{array}\right]\right)}^{\boldsymbol{\phi}_{1}^{H}}\,,\label{eq:R_yy_matrix_from}
\end{equation}
}where {\small{}$r_{0}={\displaystyle \sum_{i=0}^{v}\left|h_{i}\right|^{2}+\left(\mbox{SNR}\right)^{-1}}$,
$r_{j}=\sum_{i=j}^{v}h_{i}h_{i-j}^{*},\,\forall j\neq0$. In \cite{ourFWg},
}we showed that an upper-bound on $\mu(\boldsymbol{R}_{yy})$ in the
high SNR setting can be derived by solving the following optimization
problem 

\vspace{-1.0em}

\begin{equation}
\underset{\boldsymbol{h}}{\mbox{\mbox{max}}}\,\,\,\left|\boldsymbol{h}^{H}\boldsymbol{R}\boldsymbol{h}\right|\,\,\,\,\,\,\,\mbox{s.t.}\,\,\,\,\,\,\,\boldsymbol{h}^{H}\boldsymbol{h}=1\,,\label{eq:quadEqProb}
\end{equation}
where $\boldsymbol{h}=\left[\begin{array}{cccc}
h_{0} & h_{1} & \ldots & h_{v}\end{array}\right]^{H}$ is the length-$(v+1)$ CIR vector and $\boldsymbol{R}$ is a matrix
that has ones along the super and sub-diagonals. It is also shown
that the solution of (\ref{eq:quadEqProb}) is the eigenvector corresponding
to the maximum eigenvalue of $\boldsymbol{R}$. The eigenvalues $\lambda_{s}$
and eigenvectors $h_{j}^{(s)}$ of the matrix $\boldsymbol{R}$ have
the following simple closed-forms \cite{eigValueVector_R}:

\vspace{-1.0em}

{\small{}
\begin{eqnarray}
\lambda_{s} & = & 2\,\mbox{cos}(\frac{\pi s}{v+2})\,\,\,\,,\,\,\,\,h_{j}^{(s)}=\sqrt{\frac{2}{v+2}}\mbox{sin\ensuremath{(\frac{j\pi s}{v+2})\,,\,}}\label{eq:worst-taps}
\end{eqnarray}
}where $s,j=1,\ldots,v+1.$ Finally, by numerically evaluating $h_{j}^{(s)}$
for the maximum $\lambda_{s}$, we find that the worst-case coherence
of $\boldsymbol{R}_{yy}$ (for any $v$) is sufficiently less than
1. This observation points to the likely success of OMP in providing
the sparsest solution $\widehat{\boldsymbol{w}}_{s}$ which corresponds
to the dictionary that has the smallest coherence. Next, we will report
the results of our numerical experiments to evaluate the performance
of our proposed framework under different sparsifying dictionaries. 

\vspace{-1em}

\section{Simulation Results \label{sec:Simulation-Results}}

The CIRs used in our numerical results are unit-energy symbol-spaced
FIR filters with $v$ taps generated as zero-mean uncorrelated complex
Gaussian random variables. The CIR taps are assumed to have a uniform
power-delay-profile\footnote{This type of CIRs can be considered as a wrost-case assumption since
the inherent sparsity of other channel models, e.g., \cite{ITU-A}
and \cite{TG3C}, can lead to further shortening of the dealy spread
of the CIR. } (UPDP). The performance results are calculated by averaging over
5000 channel realizations. We use the notation $\boldsymbol{D}(\boldsymbol{\chi}_{t},\,\boldsymbol{\chi}_{e})$
to refer to a TIR designed based on the sparsifying dictionary $\boldsymbol{\chi}_{t}$
and a CSE designed based on the sparsifying dictionary $\boldsymbol{\chi}_{e}$. 

\begin{figure}
\vspace{-1.5em}

\includegraphics[scale=0.3]{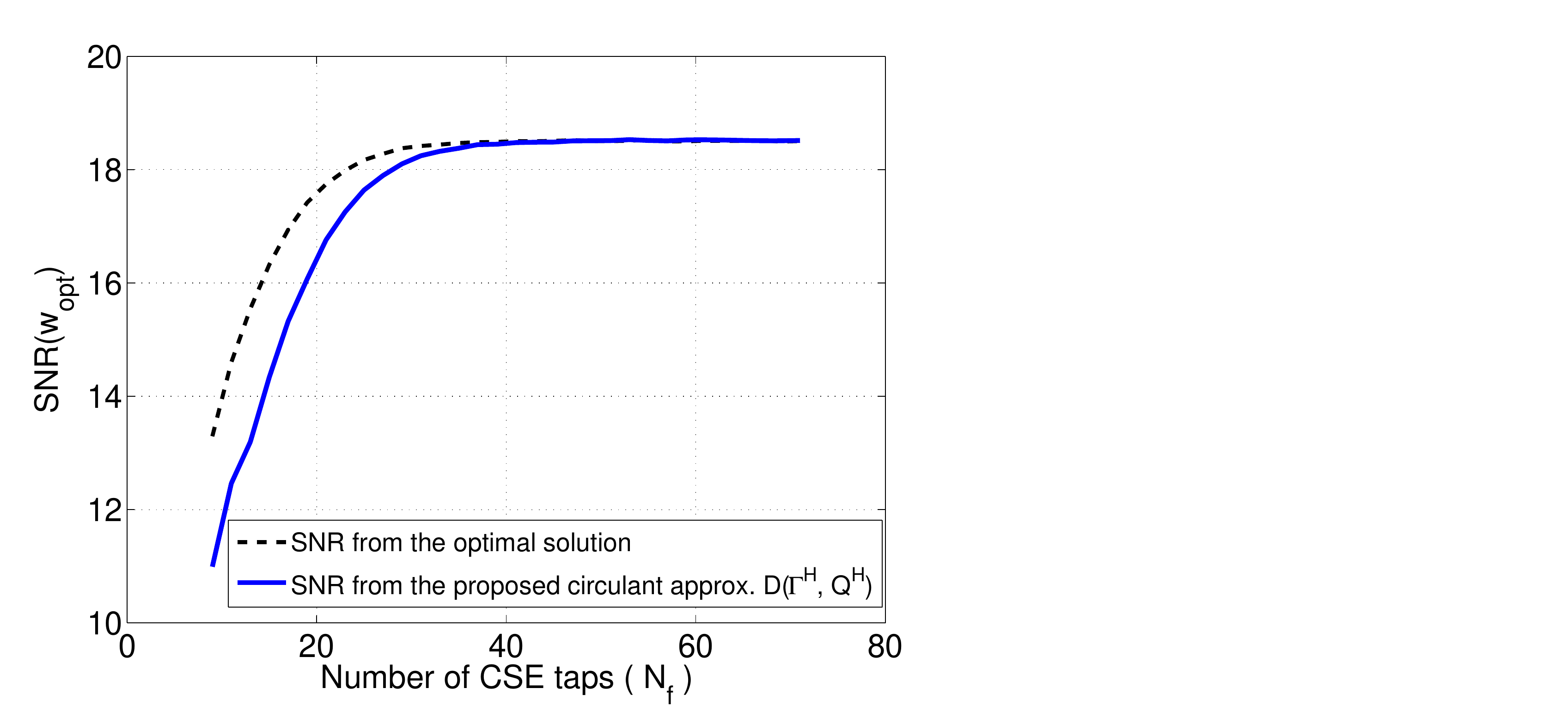}

\vspace{-1.5em}

\protect\caption{{\footnotesize{}Performance of circulant approximation based approach
for UPDP channel with $v=5$ and input $\mbox{SNR}=20\,\mbox{dB}$.}\label{fig:snr_versus_nf}}

\vspace{-1.5em}
\end{figure}

To quantify the accuracy of approximating the matrices $\boldsymbol{R}_{yy},\,\boldsymbol{R}_{\delta}$
by their equivalent circulant matrices $\boldsymbol{\overline{R}}_{yy}$
and $\overline{\boldsymbol{R}}_{\delta}$, respectively, we plot the
optimal shortening SNR and the shortening SNR obtained from the circulant
approximation versus the number of CSE taps ($N_{f}$) in Figure \ref{fig:snr_versus_nf}.
The gap between the optimal shortening SNR and the shortening SNR
from the circulant approximation approaches zero as the number of
the CSE taps increases, as expected. A good choice for $N_{f}$, to
obtain an accurate approximation, would be $N_{f}\geq5v$. 
\begin{figure}[t]
$\,\,\,$\includegraphics[scale=0.25]{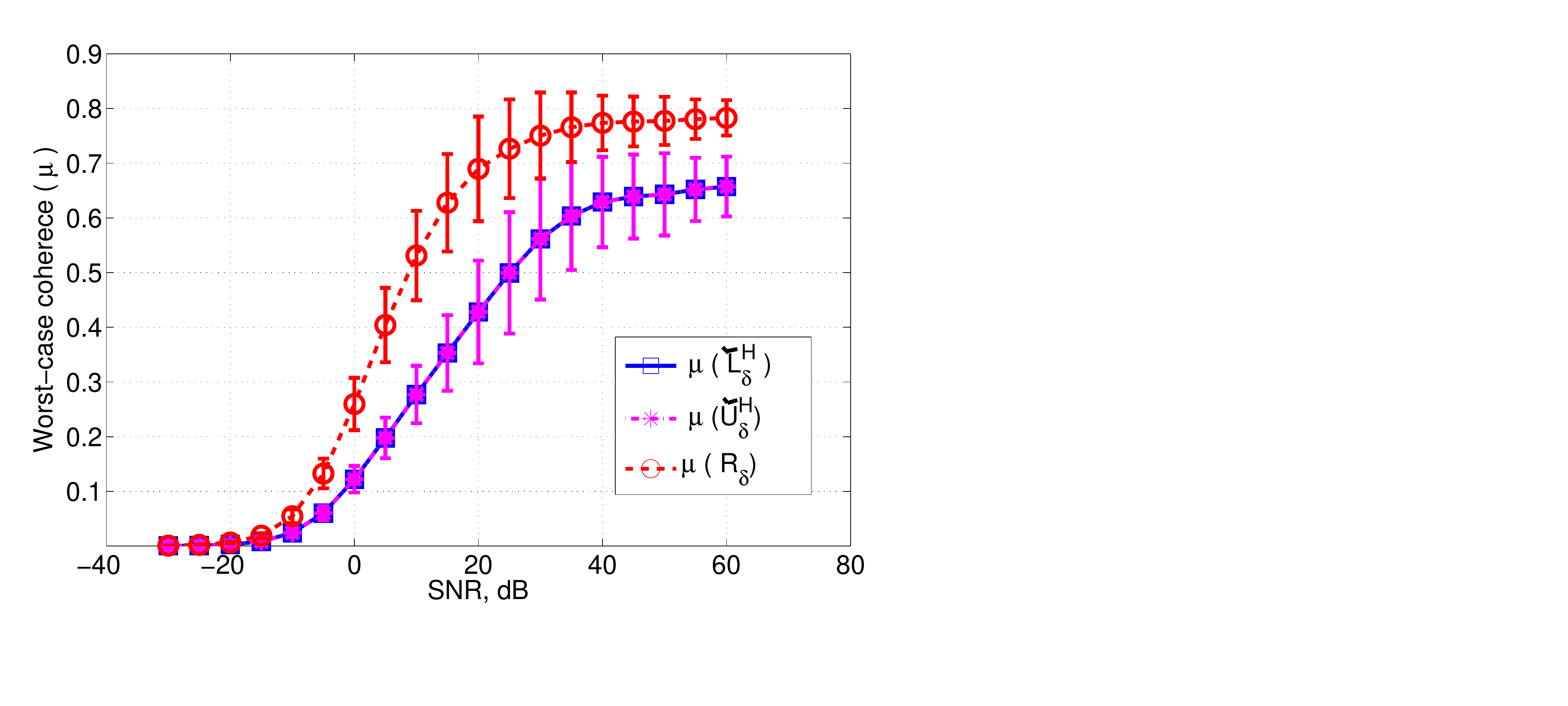}

\vspace{-3em}

\protect\caption{{\footnotesize{}Worst-case coherence for the sparsifying dictionaries
$\widetilde{\boldsymbol{L}}_{\delta}$ and $\widetilde{\boldsymbol{U}}_{\delta}$
versus input SNR for UPDP with $v=8$ and $N_{f}=80$.}\textcolor{black}{\footnotesize{}
Note that we estimate $\mu(\boldsymbol{\varPhi})$ after removing
the column corresponding to the $i^{th}$ unit tap as discussed in
(\ref{eq:quad_mse_TIR_b}).}\textcolor{black}{{} Moreover, changing
the location of the $i^{th}$ column has insignificant effect on }\textcolor{black}{\footnotesize{}$\mu(\boldsymbol{\varPhi})$.}\textcolor{black}{{}
\label{fig:coherene }}}

\vspace{-2em}
\end{figure}
\begin{figure}[t]
\vspace{-1.5em}

$\qquad$\includegraphics[scale=0.3]{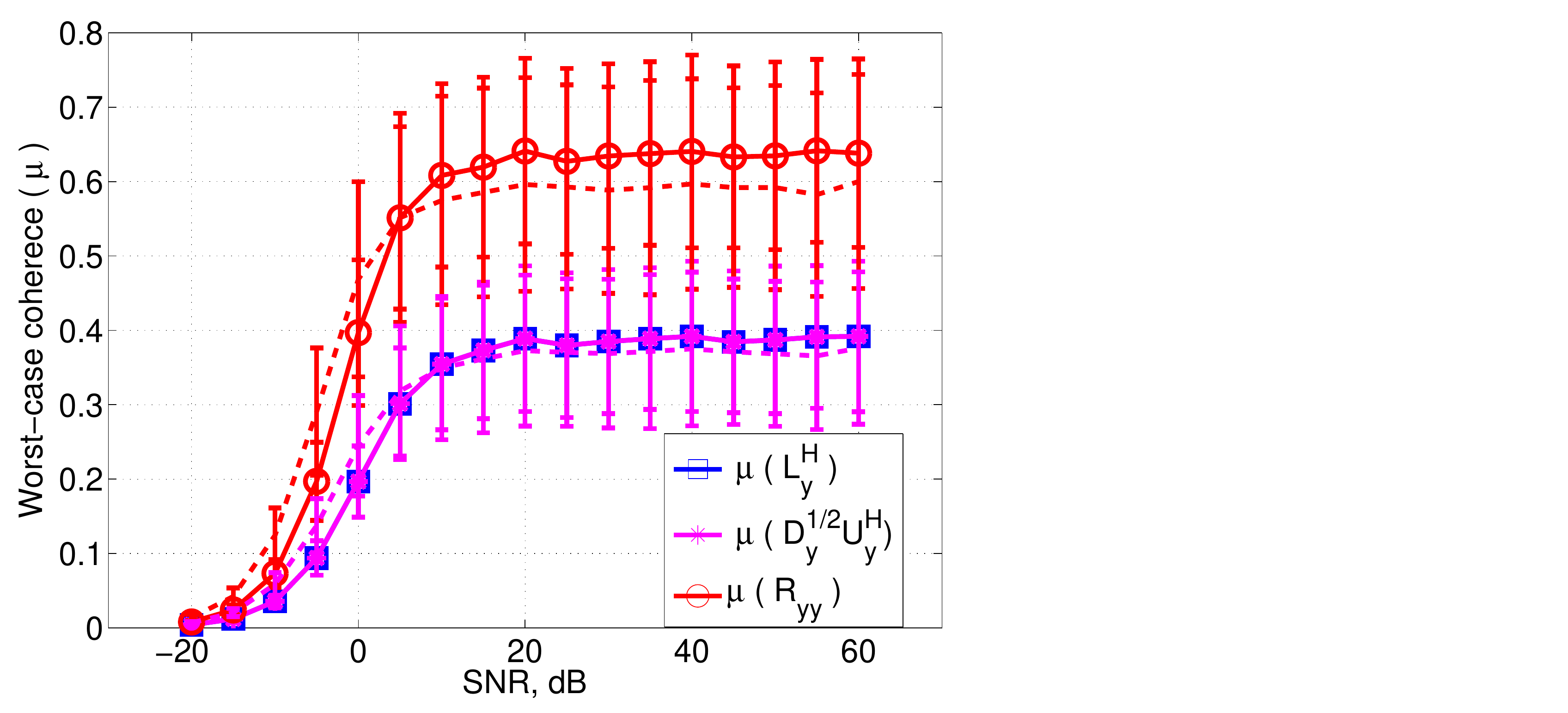}

\vspace{-1.5em}

\protect\caption{{\footnotesize{}Worst-case coherence for the sparsifying dictionaries
$\boldsymbol{L}_{y}^{H}$, $\boldsymbol{D}_{y}^{\frac{1}{2}}\boldsymbol{U}_{y}^{H}$
and $\boldsymbol{R}_{yy}$ versus input SNR for UPDP with $v=8$ and
$N_{f}=80$. Dashed lines represent the coherence of }\textcolor{black}{\footnotesize{}the
corresponding circulant approximation for $\boldsymbol{D}_{y}^{\frac{1}{2}}\boldsymbol{U}_{y}^{H}$
(i.e., $\boldsymbol{Q}^{H}$) and }{\footnotesize{}$\boldsymbol{R}_{yy}$}\textcolor{black}{\footnotesize{}
(i.e., $\boldsymbol{\overline{R}}_{yy}=$$\boldsymbol{Q}$$\boldsymbol{Q}^{H}$).}\textcolor{black}{\label{fig:cohereneR_yy}}}

\vspace{-1.0em}
\end{figure}
\begin{figure}
\includegraphics[scale=0.3]{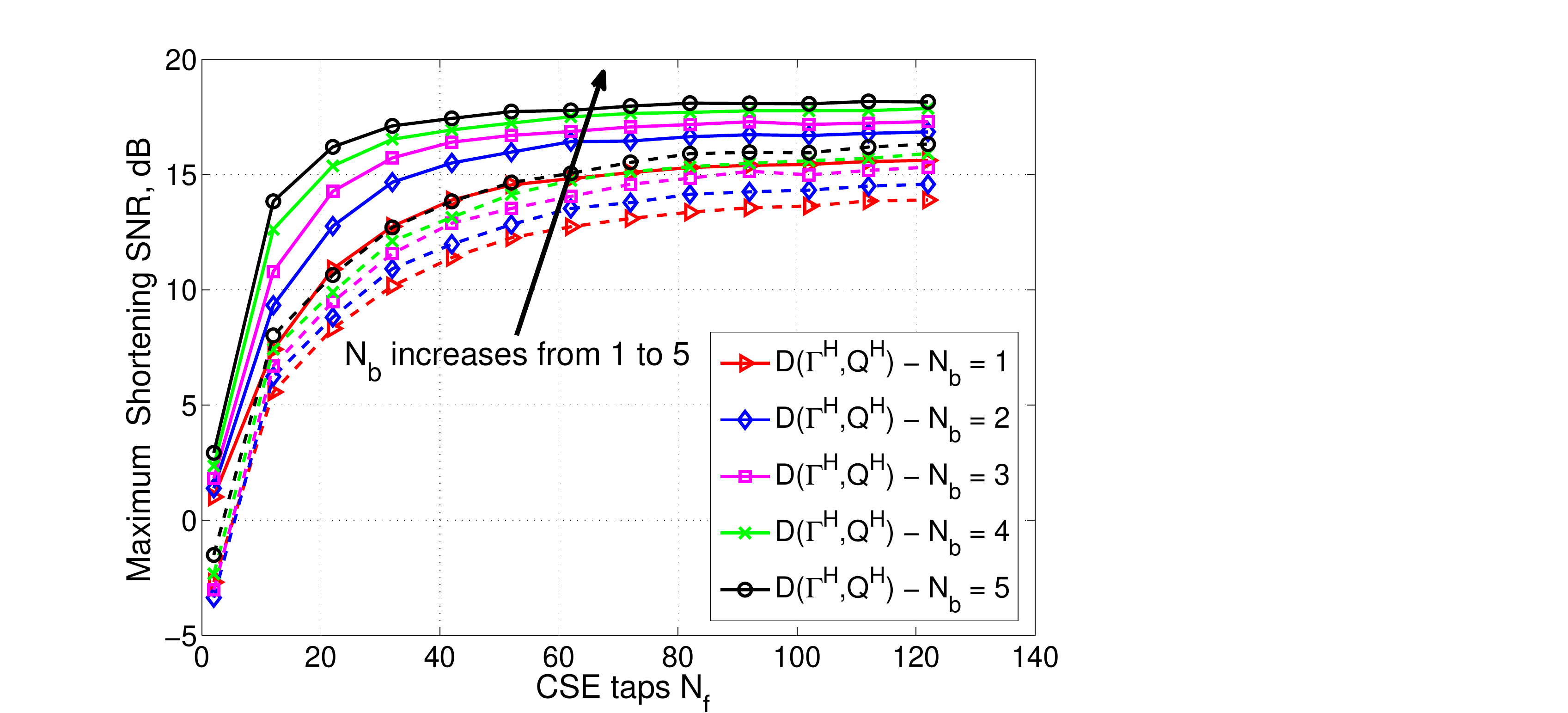}

\vspace{-1.0em}

\protect\caption{\textcolor{black}{\footnotesize{}Maximum shortening SNR versus CSE
taps for UPDP CIR with }{\footnotesize{}$v=5$, SNR = $20$ dB}\textcolor{black}{\footnotesize{}.
Solid lines represent the }{\footnotesize{}$\boldsymbol{D}(\boldsymbol{\varGamma}^{H},\boldsymbol{Q}^{H})$
}\textcolor{black}{\footnotesize{}approach while the dashed lines
represent the }{\footnotesize{}``significant-taps'' approach proposed
in} \cite{sigTaps}.\textcolor{black}{{} \label{fig:unitTap-1}}}

\vspace{-1.5em}
\end{figure}

To investigate the performance of the sparsifying dictionaries used
in our analysis in terms of coherence, we plot the worst-case coherence
versus the input SNR in Figure \ref{fig:coherene } for sparsifying
dictionaries $\widetilde{\boldsymbol{L}}_{\delta}$ and $\widetilde{\boldsymbol{U}}_{\delta}$
(which is formed by all columns of $\boldsymbol{D}_{\delta}^{\frac{1}{2}}\boldsymbol{U}_{\delta}^{H}$
except the $i^{th}$ column) generated from $\boldsymbol{R}_{\delta}$.
Note that a smaller value of $\mu(\boldsymbol{\Phi})$ indicates that
a reliable sparse approximation is more likely. Both sparsifying dictionaries
have the same $\mu\left(\boldsymbol{\varPhi}\right)$, which is strictly
less than $1$. Similarly, in Figure \ref{fig:cohereneR_yy}, we plot
the worst-case coherence of the proposed sparsifying dictionaries
which we used to design the sparse CSE. At high SNR levels, the noise
effects are negligible and, hence, the sparsifying dictionaries (e.g.,
{\small{}$\boldsymbol{R}_{yy}\approx\boldsymbol{H}\boldsymbol{H}^{H}$})
do not depend on the SNR. As a result, the coherence converges to
a constant. On the contrary, at low SNR, the noise effects dominate
the channel effects. Hence, the channel can be approximated as a memoryless
(i.e., 1 tap) channel. As such the dictionaries (e.g., {\small{}$\boldsymbol{R}_{yy}\approx\frac{1}{SNR}\boldsymbol{I}$})
can be approximated as a multiple of the identity matrix, i.e., $\mu\left(\boldsymbol{\varPhi}\right)\rightarrow0$.

In Figure \ref{fig:unitTap-1}, we compare our proposed sparse TIR
design with that in \cite{sigTaps}, which we refer to it as the ``significant-taps''
approach, in terms of shortening SNR where we plot the shortening
SNR versus CSE taps $N_{f}$ for the UPDP channel. We vary $N_{b}$,
the number of TIR taps, from $1$ (lower curve) to $5$ (upper curve).
The shortening  SNR increases as $N_{b}$ increases for all TIR designs,
as expected, and our sparse TIR outperforms, for all scenarios, the
proposed approach in \cite{sigTaps}. Notice that as $N_{b}$ increases,
the sparse TIR becomes more accurate in approximating the actual CIR. 

Next, we compare the sparse TIR designs to study the effect of $\mu\left(\boldsymbol{\varPhi}\right)$
on their performance. The OMP algorithm is used to compute the sparse
approximations. The OMP stopping criterion is set to be either a predefined
number of nonzero taps $N_{b}$ or a function of the PRE such that:
{\small{}Performance Loss ($\eta$)$=10\,\mbox{Log}_{10}\left(\frac{SNR(\boldsymbol{w}_{s})}{SNR(\boldsymbol{w}_{opt})}\right)\leq10\,\mbox{Log}_{10}\left(1+\frac{_{\delta_{eq}}}{\xi_{m}}\right)\triangleq\eta_{max}$}.
Here, $\delta_{eq}$ is computed based on an acceptable $\eta_{max}$
and, then, the coefficients of $\widehat{\boldsymbol{w}}_{s}$ are
computed using (\ref{eq:propFW}). The percentage of the active taps
is calculated as the ratio between the number of nonzero taps to the
total number of filter taps, i.e., $N_{f}$. For the optimal CSE,
where none of the coefficients is zero, the number of active filter
taps is equal to the filter span. The decision delay $\Delta$ and
the unit-tap index $i$ should be optimized to avoid performance degradation.
However, a near optimum performance is achieved by choosing these
parameters to be around $(N_{f}+v)/2$ \cite{effcompRed}.
\begin{figure}[t]
\vspace{-2em}

$\,\,\,\,\,\,\,\,\,\,$\includegraphics[scale=0.23]{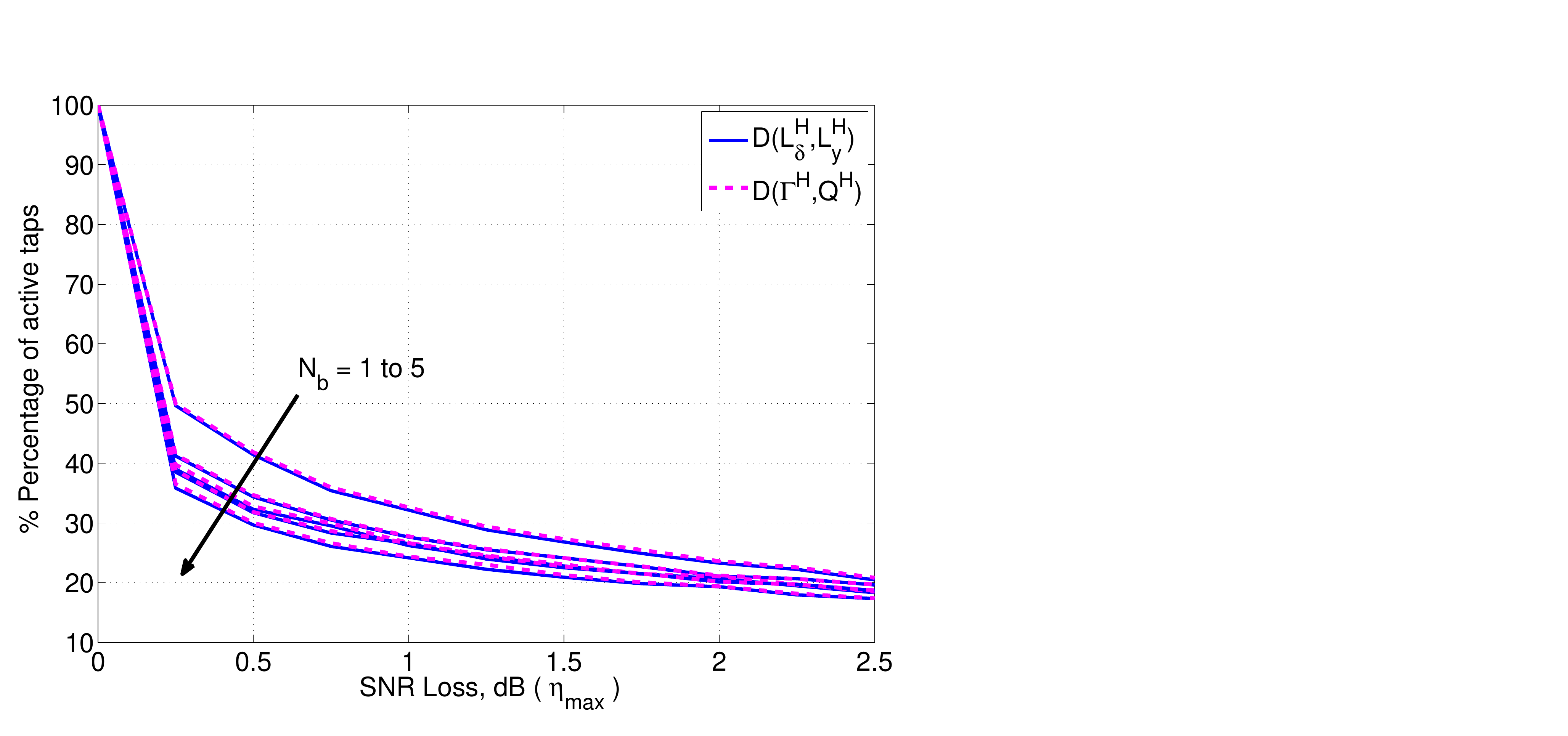}

\vspace{-1.5em}

\protect\caption{\textcolor{black}{\footnotesize{}Percentage of active CSE taps versus
the performance loss ($\eta_{max}$) for sparse TIR designs with SNR
= 20 dB, $v=5$}\textcolor{black}{{} and $N_{f}=40$.\label{fig:activeTaps_versus_eta_max_cse}}}

\vspace{-1.0em}
\end{figure}
\begin{figure}[t]
\includegraphics[scale=0.25]{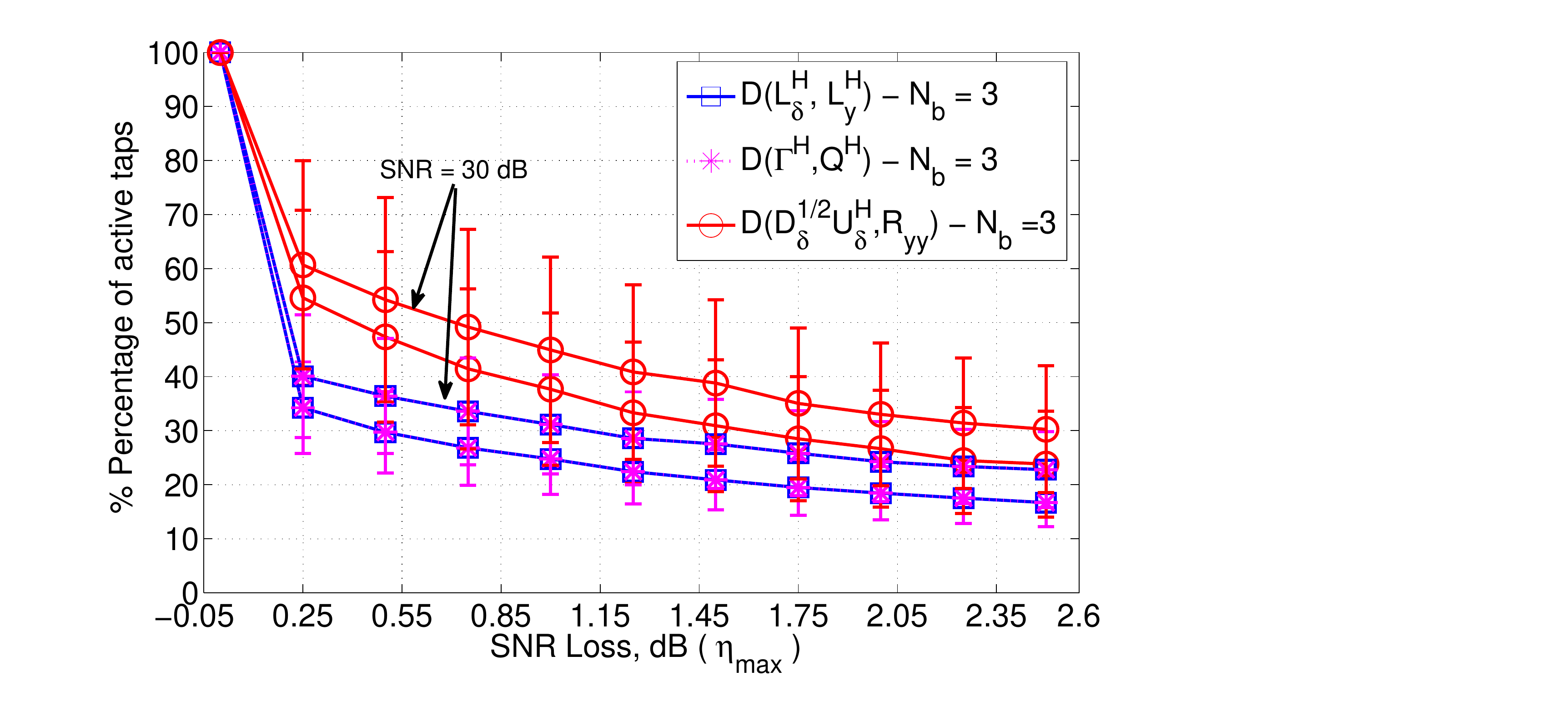}

\vspace{-1.5em}

\protect\caption{\textcolor{black}{\footnotesize{}Percentage of active CSE taps versus
the performance loss ($\eta_{max}$) for a sparse TIR with $N_{b}=3$,
SNR = 20 dB, 30 dB, $v=8$}\textcolor{black}{{} and $N_{f}=80$.\label{fig:activeTaps_versus_eta_max_cse-1}}}

\vspace{-1.5em}
\end{figure}

The effect of our sparse CSE and TIR FIR filter designs on the performance
is shown in Figure \ref{fig:activeTaps_versus_eta_max_cse}. We plot
the number of active (non-zero) CSE taps as a percentage of the total
CSE span $N_{f}$ versus the maximum loss in the shortening  SNR.
Allowing a higher loss in the shortening SNR yields a bigger reduction
in the number of CSE taps. Moreover, the active CSE taps percentage
increases as $N_{b}$ decreases because the equalizer needs more taps
to shorten the CIR to a shorter TIR. We also observe that allowing
a maximum of only 0.25 dB in SNR loss with $N_{b}=2$ results in a
substantial $60\%$ reduction in the number of CSE active taps (the
equalizer can shorten the channel using only 16 out of 40 taps). 

In Figure \ref{fig:activeTaps_versus_eta_max_cse-1}, we plot the
percentage of the active CSE taps versus $\eta_{max}$ for different
sparsifying dictionaries assuming $N_{b}=3$. We notice that a lower
active taps percentage is obtained when the coherence of the sparsifying
dictionary is smaller. For instance, allowing for $0.25$ dB SNR loss
results in a significant reduction in the number of active CSE taps.
Almost $60\%$ of the taps are eliminated when using $\boldsymbol{D}(\boldsymbol{\varGamma}^{H},\boldsymbol{Q}^{H})$
and $\boldsymbol{D}(\boldsymbol{L}_{\delta}^{H},\,\boldsymbol{L}_{y}^{H})$
at SNR equals to 30 dB. $\boldsymbol{D}(\boldsymbol{D}_{\delta}^{\frac{1}{2}}\boldsymbol{U}_{\delta}^{H},\,\boldsymbol{R}_{yy})$
needs more active taps to achieve the same SNR loss as that of the
other dictionaries due to its higher coherence for a given constraint
on TIR taps $N_{b}$. This suggests that the smaller the worst-case
coherence is, the sparser the CSE will be. Moreover, a lower sparsity
level (active taps percentage) is achieved at higher SNR levels which
is consistent with the previous findings (e.g., in \cite{tapPositions07}).
Moreover, reducing the number of active taps decreases the design
complexity and, consequently, the power consumption since a smaller
number of complex multiply-and-add operations are required.

\vspace{-0.5em}

\section{Conclusions\label{sec:Conclusion-and-Future}}

In this paper, we proposed a general framework for sparse CSE and
TIR FIR designs based on a sparse approximation formulation using
different dictionaries. Based on the asymptotic equivalence of Toeplitz
and circulant matrices, we proposed reduced-complexity designs, for
both CSE and TIR FIR filters, where matrix factorizations can be carried
out efficiently using the FFT and inverse FFT with negligible performance
loss as the number of filter taps increases. In addition, we analyzed
the coherence of the proposed dictionaries involved in our design
and showed that the dictionary with the smallest coherence gives the
sparsest filter design. The significance of our approach was also
quantified through simulations. 

\vspace{-1.0em}

\bibliographystyle{IEEEtran}
\bibliography{icc16Ref}

\end{document}